\documentclass[twocolumn,aps,showpacs,prl,superscriptaddress]{revtex4} 

\usepackage{sidecap}
\usepackage{ulem}
\usepackage{epsfig}
\usepackage{amsmath,amssymb,amsthm}
\usepackage{graphicx}
\usepackage{bm}
\usepackage{color}

\DeclareMathAlphabet{\mathpzc}{OT1}{pzc}{m}{it}

\begin{document}

\renewcommand{\textfraction}{0.00}


\newcommand{\vAi}{{\cal A}_{i_1\cdots i_n}} 
\newcommand{\vAim}{{\cal A}_{i_1\cdots i_{n-1}}} 
\newcommand{\vAbi}{\bar{\cal A}^{i_1\cdots i_n}}
\newcommand{\vAbim}{\bar{\cal A}^{i_1\cdots i_{n-1}}}
\newcommand{\htS}{\hat{S}} 
\newcommand{\htR}{\hat{R}}
\newcommand{\htB}{\hat{B}} 
\newcommand{\htD}{\hat{D}}
\newcommand{\htV}{\hat{V}} 
\newcommand{\cT}{{\cal T}} 
\newcommand{\cM}{{\cal M}} 
\newcommand{\cMs}{{\cal M}^*}
\newcommand{\vk}{\vec{\mathbf{k}}}
\newcommand{\bk}{\bm{k}}
\newcommand{\kt}{\bm{k}_\perp}
\newcommand{\kp}{k_\perp}
\newcommand{\km}{k_\mathrm{max}}
\newcommand{\vl}{\vec{\mathbf{l}}}
\newcommand{\bl}{\bm{l}}
\newcommand{\bK}{\bm{K}} 
\newcommand{\bb}{\bm{b}} 
\newcommand{\qm}{q_\mathrm{max}}
\newcommand{\vp}{\vec{\mathbf{p}}}
\newcommand{\bp}{\bm{p}} 
\newcommand{\vq}{\vec{\mathbf{q}}}
\newcommand{\bq}{\bm{q}} 
\newcommand{\qt}{\bm{q}_\perp}
\newcommand{\qp}{q_\perp}
\newcommand{\bQ}{\bm{Q}}
\newcommand{\vx}{\vec{\mathbf{x}}}
\newcommand{\bx}{\bm{x}}
\newcommand{\tr}{{{\rm Tr\,}}} 
\newcommand{\bc}{\textcolor{blue}}

\newcommand{\beq}{\begin{equation}}
\newcommand{\eeq}[1]{\label{#1} \end{equation}} 
\newcommand{\ee}{\end{equation}}
\newcommand{\bea}{\begin{eqnarray}} 
\newcommand{\eea}{\end{eqnarray}}
\newcommand{\beqar}{\begin{eqnarray}} 
\newcommand{\eeqar}[1]{\label{#1}\end{eqnarray}}
 
\newcommand{\half}{{\textstyle\frac{1}{2}}} 
\newcommand{\ben}{\begin{enumerate}} 
\newcommand{\een}{\end{enumerate}}
\newcommand{\bit}{\begin{itemize}} 
\newcommand{\eit}{\end{itemize}}
\newcommand{\ec}{\end{center}}
\newcommand{\bra}[1]{\langle {#1}|}
\newcommand{\ket}[1]{|{#1}\rangle}
\newcommand{\norm}[2]{\langle{#1}|{#2}\rangle}
\newcommand{\brac}[3]{\langle{#1}|{#2}|{#3}\rangle} 
\newcommand{\hilb}{{\cal H}} 
\newcommand{\pleft}{\stackrel{\leftarrow}{\partial}}
\newcommand{\pright}{\stackrel{\rightarrow}{\partial}}

\title{Generalization of radiative jet energy loss to non-zero magnetic mass}
\author{Magdalena Djordjevic}
\affiliation{Institute of Physics Belgrade, University of Belgrade, Serbia}
\author{Marko Djordjevic}
\affiliation{Faculty of Biology, University of Belgrade, Serbia}

\begin{abstract}
Reliable predictions for jet quenching in ultra-relativistic heavy ion 
collisions require  accurate computation of radiative energy loss. With this 
goal, an energy loss formalism in a realistic finite size dynamical QCD medium 
was recently developed. While this formalism assumes zero magnetic mass - in 
accordance with the one-loop perturbative calculations - different 
non-perturbative approaches report a non-zero magnetic mass at RHIC and LHC.
We here generalize the energy loss to consistently include a possibility 
for existence of non-zero magnetic screening. We also present how the inclusion
of finite magnetic mass changes the energy loss results. Our analysis indicates 
a fundamental constraint on magnetic to electric mass ratio.
\end{abstract}

\date{\today} 
 

\maketitle

\section{Introduction}
\label{sec1}
%
Heavy flavor suppression is considered to be a powerful tool to study the 
properties of a QCD medium created in ultra-relativistic heavy ion 
collisions~\cite{Brambilla}. The suppression results from the energy loss 
of high energy partons moving through the plasma~\cite{suppression}. Therefore, 
reliable computations of heavy quark energy loss are essential 
for the reliable predictions of jet suppression. In~\cite{MD_PRC,DH_PRL}, we 
developed a theoretical formalism for the calculation of the first order 
in opacity radiative energy loss in a dynamical QCD medium (see also a 
viewpoint in~\cite{Gyulassy_viewpoint}). That study models radiative energy 
loss in a realistic {\it finite size} QCD medium with {\it dynamical} 
constituents, therefore removing a major approximation of static scattering 
centers present in previous calculations 
(see e.g.~\cite{GLV,Gyulassy_Wang,Wiedemann,WW,DG_Ind,ASW,MD_TR}).

The dynamical energy loss formalism~\cite{MD_PRC,DH_PRL} is based on HTL 
perturbative QCD, which requires zero magnetic mass. However, different 
non-perturbative approaches~\cite{Maezawa,Nakamura,Hart,Bak} suggest a 
non-zero magnetic mass at RHIC and LHC. This, therefore, arises a question if 
finite magnetic mass can be consistently included in the dynamical energy 
loss calculations, and how this inclusion would modify the energy loss results.
%
\section{Radiative energy loss in a dynamical QCD medium} 
\label{sec2}
In~\cite{MD_PRC,DH_PRL}, we used finite temperature field theory (HTL 
approximation) and calculated the  radiative energy loss in a finite size 
dynamical QCD medium. The obtained expression for the energy loss is given 
by Eq.~(\ref{DeltaEDyn}): 
%
\beqar
\frac{\Delta E_{\mathrm{rad}}}{E} 
= \frac{C_R \alpha_s}{\pi}\,\frac{L}{\lambda_\mathrm{dyn}}  
    \int dx \,\frac{d^2k}{\pi} \,\frac{d^2q}{\pi} \, v(\bq) \, f(\bk,\bq,x) ,
\eeqar{DeltaEDyn}
%
where

\beqar
f(\bk,\bq,x)&=& 2\, \left(1-\frac{\sin{\frac{(\bk{+}\bq)^2+\chi}{x E^+} \, L}} 
    {\frac{(\bk{+}\bq)^2+\chi}{x E^+}\, L}\right) \nonumber \\
&&\hspace*{-0.7cm} \times \frac{(\bk{+}\bq)}{(\bk{+}\bq)^2+\chi}
    \left(\frac{(\bk{+}\bq)}{(\bk{+}\bq)^2+\chi}
    - \frac{\bk}{\bk^2+\chi}
    \right).
\eeqar{fkqx}

In  Eqs.~(\ref{DeltaEDyn}) and~(\ref{fkqx}), $L$ is the length of the finite 
size dynamical QCD medium and $E$ is the jet energy. $\bk$ is transverse 
momentum of radiated gluon, while $\bq$ is transverse momentum of the 
exchanged (virtual) gluon.  $\alpha_s = \frac{g^2}{4 \pi}$ is coupling 
constant and $C_R=\frac{4}{3}$. $v(\bq)$ is the effective crossection in 
dynamical QCD medium and $\lambda_\mathrm{dyn}^{-1} \equiv C_2(G) \alpha_s T = 
3 \alpha_s T$ ($C_2(G)=3$) is defined as ``dynamical mean free path'' (see 
also~\cite{DH_Inf}).  $\chi\equiv M^2 x^2 + m_g^2$, where $x$ is the 
longitudinal momentum fraction of the heavy quark carried away by the emitted 
gluon, $M$ is the mass of the heavy quark, $m_g=\mu_E/\sqrt 2$ is the 
effective mass for gluons with hard momenta $k > T$~\cite{DG_TM}, and $\mu_E$ 
is the Debye mass. We assume constant coupling $g$. Furthermore, we note that 
in Eq.~(\ref{DeltaEDyn}) effective crossection $v(\bq)$ represents the 
interaction between the jet and exchanged gluon, while $f(\bk,\bq,x)$ 
represents the interaction between the jet and radiated 
gluon~\cite{MD_PRC,DH_PRL}. 

The goal of this section is to start from the above expression, and generalize 
it to include the existence of non-zero magnetic mass~\cite{Maezawa,Nakamura,Hart,Bak}. To proceed, we note that the inclusion of magnetic mass modifies the
gluon self energy, and therefore our goal is to study how modified self 
energy of radiated and exchanged gluons change the energy loss result. We also 
note that from~\cite{MD_PRC}, it is straightforward to show that
non-zero magnetic mass does not alter the factorization 
($v(\bq) \, f(\bk,\bq,x)$) in the integrand of Eq.~(\ref{DeltaEDyn}), due to 
the fact that the factorization does not depend on specific form of self 
energy. Since $v(\bq)$ depends only on the exchanged gluon self energy, 
while $f(\bk,\bq,x)$ depends only on a radiative gluon self energy, we below 
separately study how the inclusion of magnetic mass will modify $v(\bq)$ and 
$f(\bk,\bq,x)$.

\bigskip
{\bf Modification of the effective crossection due to magnetic screening. }
The effective crossection $v(\bq)$ can be written in the following form
\beqar
v(\bq) &=& v_L(\bq) - v_T(\bq) \, ,
\eeqar{vLT}
where $v_T(\bq)$ ($v_L(\bq)$) is transverse (longitudinal) contribution to the effective crossection, given by~\cite{Aurenche,MD_PRC}
\beqar
v_{T,L}(\bq) &=& \frac{1}{\bq^2+{\rm Re}\Pi_{T,L}(\infty)} - 
\frac{1}{\bq^2+{\rm Re}\Pi_{T,L}(0)} ,
\eeqar{vTL}
where $\Pi_T$ and $\Pi_L$ are gluon self energies.
While in~\cite{MD_PRC,DH_PRL} the derivation of the effective crossection was 
made through a hard thermal loop for the self-energy $\Pi$, one should note 
that the crossection does not depend on specific form of gluon self 
energy~\cite{Aurenche}. That is, the expression is valid for any self-energy 
satisfying the following assumptions~\cite{Aurenche}:
\begin{enumerate}
\item $\Pi$ depends only on $x\equiv k_0/k$
\item ${\rm Im}\,\Pi(x=0)=0$
\item ${\rm Im}\,\Pi(x)=0$ if $x\ge 1$
\item ${\rm Re}\,\Pi(x)\ge 0$ if $x\ge 1$,
\end{enumerate}
which are reasonable approximations for any system of well defined 
quasiparticles.

Therefore, we see that the result given by Eq.~(\ref{vTL}) depends only on 
four numbers: ${\rm Re}\,\Pi_{T,L}(\infty)$ and ${\rm Re}\,\Pi_{T,L}(0)$; due to 
this, we don't need to know the full gluon propagator to generalize the 
effective crossection to the case of finite magnetic screening. The first two 
numbers are the masses of the longitudinal and the transverse gluons at zero 
momentum (so called plasmon masses). These are shown to be equal due to
Slavnor-Taylor identities~\cite{Kobes,Braaten,Dirks}. Physically,
this property means that there is no way to distinguish transverse and
longitudinal modes for a particle at rest~\cite{Aurenche}. Therefore, we need 
only to introduce one plasmon mass:
\begin{equation}
{\rm Re}\,\Pi_{T}(\infty)={\rm Re}\,\Pi_{L}(\infty)\equiv \mu_{pl}^2\; .
\end{equation}
The second two quantities are squares of the screening masses
for the transverse and longitudinal static gluon exchanges. The
longitudinal (electric) screening mass is the familiar Debye mass:
\begin{equation}
\mu_E^2\equiv {\rm  Re}\,\Pi_{L}(0)
\end{equation}
In the HTL approximation, there is no screening for the transverse
static gluons, but this is not expected to hold generally. The
corresponding screening mass is the magnetic mass, and is denoted 
\begin{equation}
\mu_{M}^2\equiv {\rm Re}\,\Pi_{T}(0)\; .
\end{equation}
The general expressions for the transverse and longitudinal contributions to 
the effective crossections $v_{T,L}(\bq)$ then become 
\beqar
v_{L, T}(\bq) &=&  \frac{1}{(\bq^2+\mu_{pl}^2)} - \frac{1}{(\bq^2+\mu_{E,M}^2)} 
\eeqar{vTL_fin_mag}
After replacing the expressions for $v_{L,\, T}(\bq)$ from 
Eq.~(\ref{vTL_fin_mag}) into Eq.~(\ref{vLT}), we finally obtain the 
expression for the effective crossection in the case of non-zero magnetic mass:
\beqar
v (\bq) =  \frac{\mu_E^2-\mu_M^2}{(\bq^2+\mu_M^2) (\bq^2+\mu_E^2)}\, .
\eeqar{v_mag}

Note that $v(\bq)$ in Eq.~(\ref{v_mag}) does not depend on plasmon 
mass. In other words, all the dependence on the plasmon mass drops out in 
this expression. This seems reasonable given that $v(\bq)$ involves only 
space-like gluon exchanges (see~\cite{DH_Inf,DH_PRL,MD_PRC}), while the 
plasmon mass is a property of time-like gluons~\cite{Aurenche}. Therefore, we 
only need to know the two screening masses $\mu_E$ and $\mu_M$, in order to 
generalize the effective crossection to non-zero magnetic mass. 

\bigskip 

{\bf Modification of $f(\bk,\bq,x)$ due to magnetic screening.}
As we discussed above, the introduction of the 
magnetic mass leads to the modification of the exchanged and radiated gluon 
self energy. In this subsection, we study how the introduction of the 
magnetic mass in the radiated gluon self energy modifies the radiative 
energy loss.

To proceed with this study, we note that all radiative energy loss 
calculations~\cite{GLV,Gyulassy_Wang,Wiedemann,WW,DG_Ind,ASW,MD_TR,AMY,MD_PRC} 
are performed by assuming validity of the soft gluon ($\omega \ll E$) and soft 
rescattering ($\omega \gg |\bk| \sim |\bq| \sim q_0,q_z$) approximations. 
Within these approximations, we showed that in the finite temperature QCD 
medium radiated gluons have similar dispersion relation as in the vacuum, 
with the difference that the gluons now acquire a ``mass''~\cite{DG_TM}. We 
also showed that the gluon mass in the medium is approximately equal to the 
value of gluon self energy at $x=1$~\cite{DG_TM,Rebhan} (so called asymptotic 
mass $m_\infty=\sqrt{\Pi_T(x=1)}$). 

Therefore, analogously to the previous 
section, we see that the dependence of the $f(\bk,\bq,x)$ on gluon self energy
reduces to just a single number: $\Pi_T(x=1)$, which is defined as a square of 
gluon mass $m_g$. Due to this, instead of knowing the full gluon propagator, 
we only need to know how $m_g$ changes in order to obtain how $f(\bk,\bq,x)$ 
is modified in the case of non-zero magnetic mass.  

In principle, gluon mass may change with the introduction of non-zero magnetic 
screening, but (to our knowledge) no study up to now addressed how 
non-perturbative calculations would modify the gluon asymptotic mass. 
Consequently, our approach in the next section is to introduce an ansatz in 
order to numerically investigate how perturbations of $m_g$, for a magnitude 
corresponding to magnetic mass, change radiative energy loss results.  

\bigskip
{\bf Modification of the energy loss expression due to magnetic screening.} 
After replacing the effective crossection $v(\bq)$ (see Eq.~(\ref{v_mag})) 
into Eq.~(\ref{DeltaEDyn}), the total energy loss becomes
\beqar
\frac{\Delta E_{\mathrm{rad}}}{E} 
&=& \frac{C_R \alpha_s}{\pi}\,\frac{L}{\lambda_\mathrm{dyn}} (\mu_E^2-\mu_M^2) 
  \int dx \,\frac{d^2k}{\pi} \,\frac{d^2q}{\pi} \, \nonumber \\
&& \hspace*{0.5cm} 
  \times \frac{1}{(\bq^2+\mu_M^2) (\bq^2+\mu_E^2)}\, f(\bk,\bq,x) \, ,
\eeqar{DeltaEDyn_new}
%
where $f(\bk,\bq,x)$ is given by Eq.~(\ref{fkqx}). Note that in 
Eq.~(\ref{fkqx}), $\chi\equiv M^2 x^2 + M_g^2$, where the gluon mass
$M_g$ can now be different from $m_g=\mu_E/\sqrt{2}$ (see previous subsection). 

\bigskip
{\bf A constraint on the magnetic mass range.} We first discuss one 
interesting qualitative observation, that comes directly from 
Eq.~(\ref{DeltaEDyn_new}): Since integrand in Eq.~(\ref{DeltaEDyn_new}) is 
positive definite, if magnetic 
mass becomes larger than electric mass, the net energy loss becomes negative. 
Therefore, if magnetic mass is larger than electric mass, the quark jet 
would, overall, start to gain (instead of lose) energy in this type of plasma. 
The origin for this effect can be traced from Eq.~(\ref{vTL_fin_mag}): if the 
magnetic mass is larger than electric mass, the energy gain from magnetic 
contribution becomes so large, that it, overall, leads to the total energy gain 
of the jet. One should note that such a gain would involve transfer of energy 
of disordered motion of plasma constituents, to energy of ordered jet motion. 
Such transfer of ``low'' to ``high'' quality energy would be in a violation 
of the second law of thermodynamics. From this follows that it is impossible to 
create a plasma with magnetic mass larger than electric, which places a 
fundamental limit on magnetic mass range. Indeed, in an agreement with this 
limit, various non-perturbative approaches~\cite{Maezawa,Nakamura,Hart,Bak} 
suggest that, at RHIC and LHC, $0.4 < \mu_M/\mu_E < 0.6$. 

\section{Numerical results} 
In this section, we numerically study how the inclusion of non-zero magnetic 
mass modifies the energy loss results. To address this, we consider a 
quark-gluon plasma of temperature $T{\,=\,}225$\,MeV, with $n_f{\,=\,}2.5$ 
effective light quark flavors and strong interaction strength 
$\alpha_s{\,=\,}0.3$, as representative of average conditions encountered 
in Au+Au collisions at RHIC. For the light quark jets we assume that their 
mass is dominated by the thermal mass $M{\,=\,}\mu/\sqrt{6}$, where 
$\mu{\,=\,}gT\sqrt{1{+}N_f/6}\approx 0.5$ GeV is the Debye screening 
mass. The charm mass is taken to be $M{\,=\,}1.2$\,GeV, and for 
the bottom mass we use $M{\,=\,}4.75$\,GeV.  To simulate (average) 
conditions in Pb+Pb collisions at the LHC, we use the temperature of 
the medium of $T{\,=\,}400$\,MeV.
%
\begin{SCfigure*}
\epsfig{file=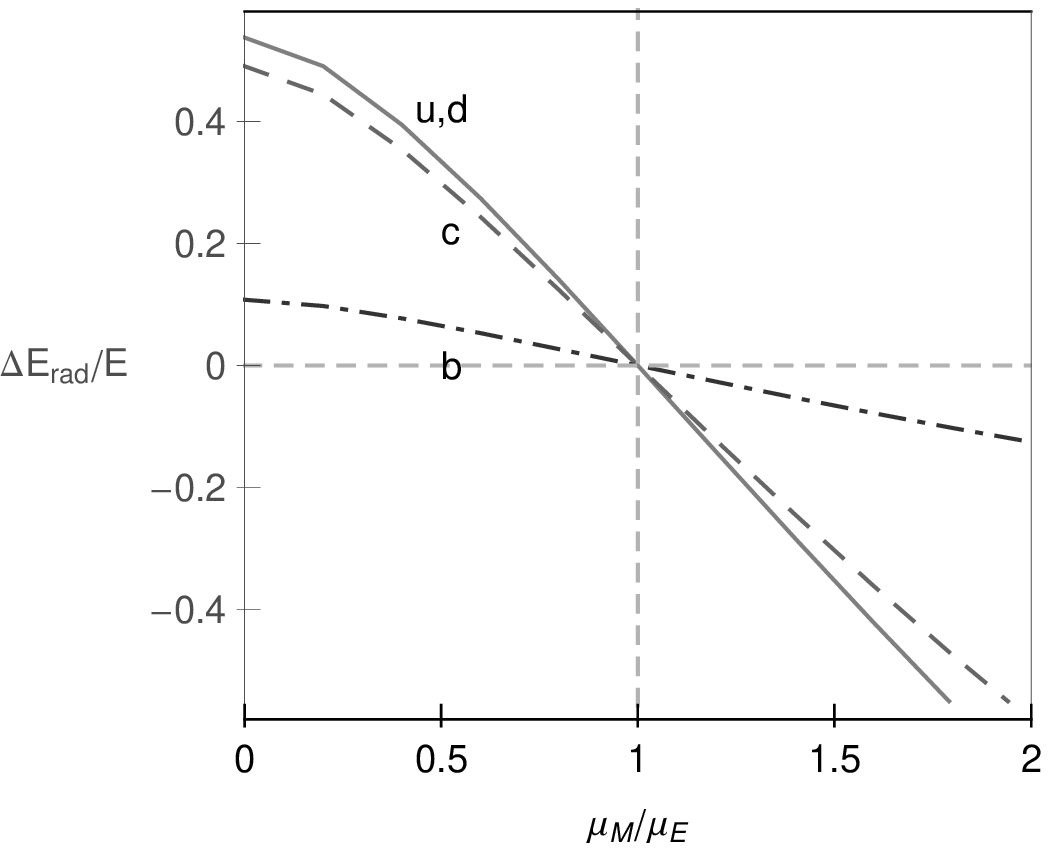,width=2.3in,height=1.7in,clip=5,angle=0}
\hspace{0.1cm}
\epsfig{file=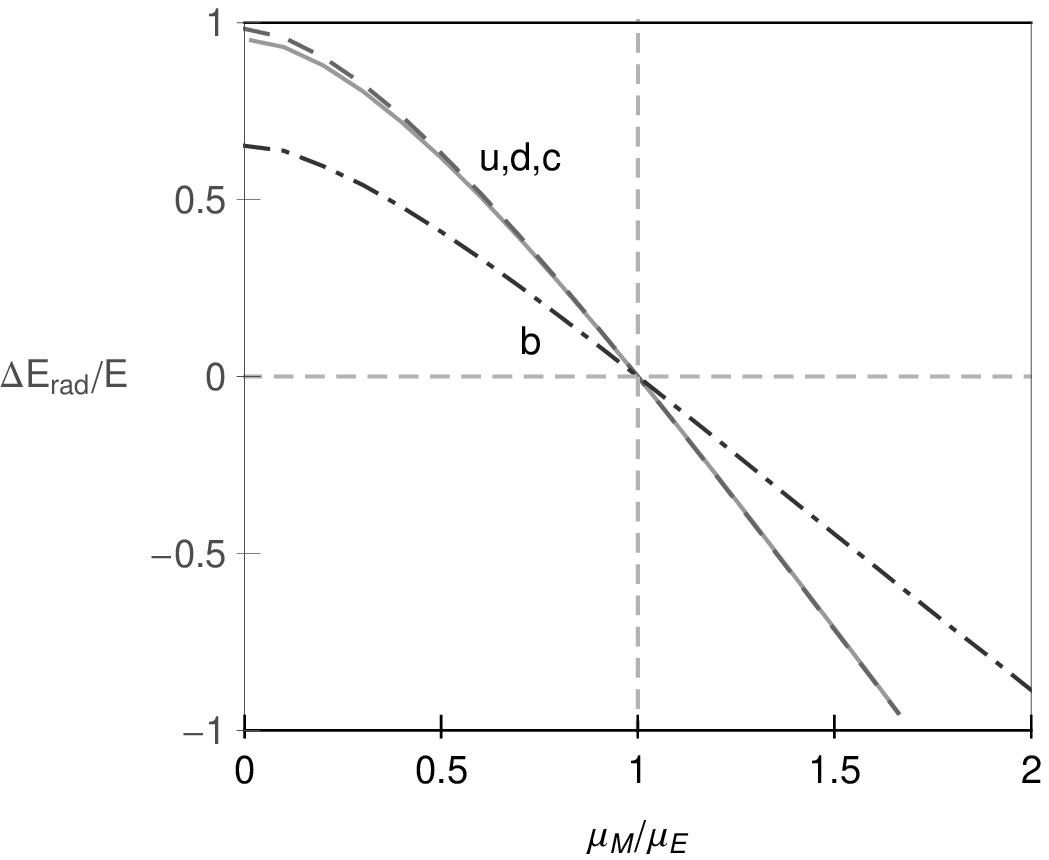,width=2.3in,height=1.7in,clip=5,angle=0}  
\hspace{0.05cm}
\caption{Fractional radiative energy loss is shown as a function of magnetic 
and electric mass ratio. Assumed path length is $L=5$\,fm and initial jet 
energy is 10 (50)~GeV for a left (right) panel. Full, dashed and dot-dashed 
curves correspond to light, charm and bottom quark respectively. 
Note that for left (right) panel, we assume RHIC (LHC) conditions, with a 
medium of temperature $T=225$ (400) MeV.}
\label{MagnDep}
\end{SCfigure*}

\begin{SCfigure*}
\epsfig{file=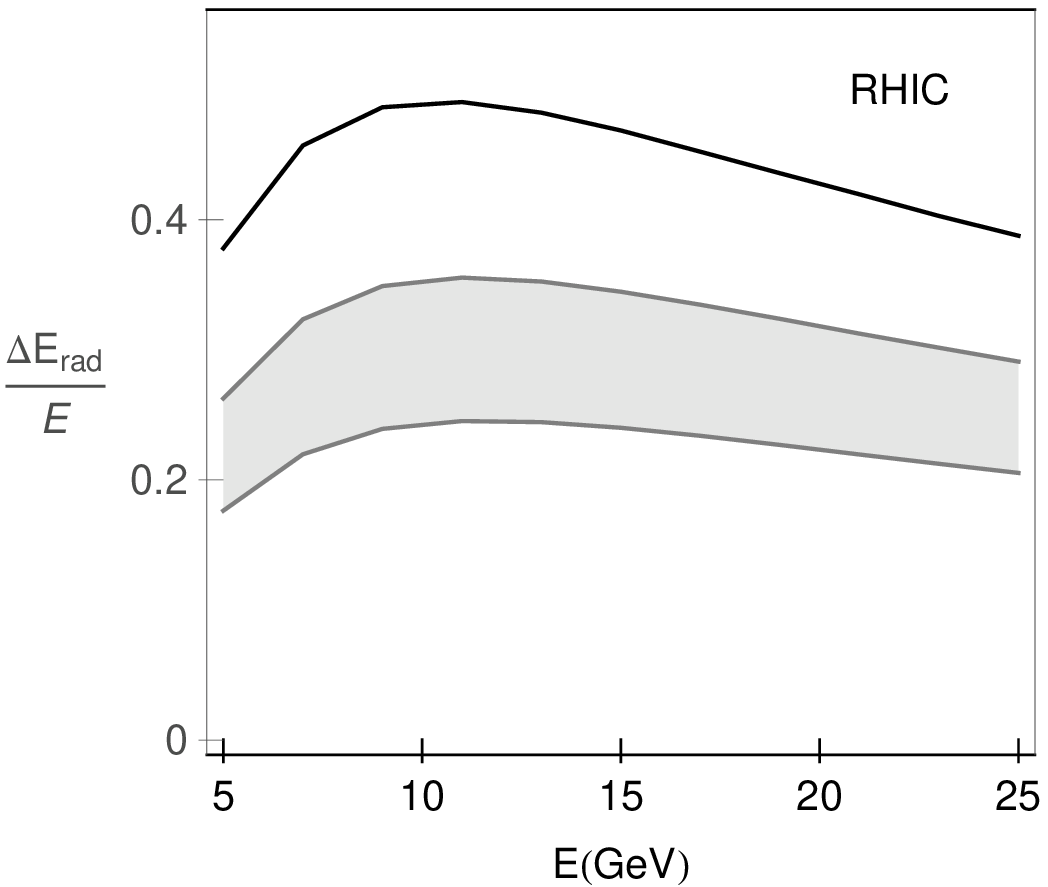,width=2.in,height=1.6in,clip=5,angle=0}
\hspace{0.1cm}
\epsfig{file=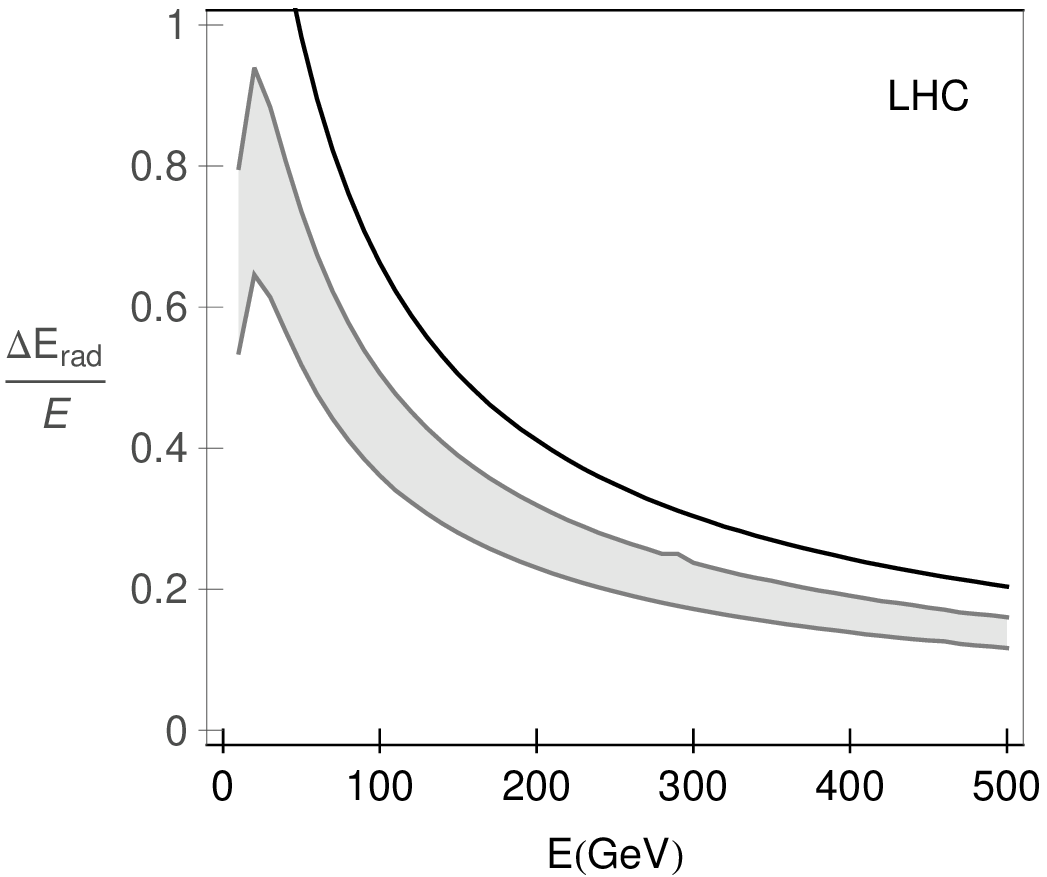,width=2.in,height=1.6in,clip=5,angle=0}  
\hspace{0.05cm}
\caption{Fractional radiative energy loss for an assumed path length 
$L=5$\,fm as a function of momentum for charm quarks. Left (right) panel corresponds to RHIC (LHC) conditions. Full curve corresponds to the case when magnetic mass is zero.  Gray band corresponds to the energy loss when magnetic mass is non-zero (i.e.  $0.4 < \mu_M/\mu_E < 0.6$). Upper (lower) boundary of the band corresponds to the case $\mu_M/\mu_E=0.4$ 
($\mu_M/\mu_E =0.6$). }
\label{DynEdep}
\end{SCfigure*}

We first investigate how possible changes of gluon mass (i.e. $f(\bk,\bq,x)$)
due to non-zero magnetic screening may change radiative energy loss (see the 
previous section). To investigate this, we introduce an ansatz that both 
electric and magnetic masses equally contribute to gluon self energy at $x=1$ 
(i.e. $M_g=\sqrt{\frac{\mu_E^2+\mu_M^2}{2}}$). With this ansatz, which changes 
the gluon mass for a magnitude comparable to magnetic mass, we obtain a 
negligible change in radiative energy loss (data not shown). For simplicity, 
we will therefore further assume that the gluon mass of radiated gluon remains 
the same as in~\cite{MD_PRC,DG_TM}, i.e. that $M_g=\mu_E/\sqrt{2}$. 
Consequently, in the rest of this section, we numerically study how the 
inclusion of magnetic mass into the effective crossection modifies the energy 
loss results compared to the results presented in~\cite{MD_PRC}.

Energy loss dependence on the magnetic mass is shown in Fig.~\ref{MagnDep} 
for RHIC and LHC case. As expected, we see that energy loss decreases with the 
increase in magnetic mass. Note that, when magnetic masses becomes larger 
than electric mass, the net energy loss becomes negative, as discussed in the 
previous section. In Fig.~\ref{DynEdep}, we show momentum dependence of 
fractional energy loss, where we concentrate on the range 
$0.4 < \mu_M/\mu_E < 0.6$, as suggested by various non-perturbative 
approaches~\cite{Maezawa,Nakamura,Hart,Bak}. We see that finite magnetic mass 
reduces the energy loss in dynamical QCD medium by 25\% to 50\%. 

We note that, contrary to what one may naively expect, majority of the energy 
loss decrease does not come from the introduction of magnetic screening in 
the denominator of the effective crossection (see Eq.~(\ref{v_mag})). In fact, 
the major decrease in the energy loss actually comes from the presence of the 
magnetic mass in the numerator of the energy loss expression (numerical 
results not shown). For example, for the ratio $\mu_M/\mu_E=0.5$, 25\% 
decrease in the energy loss comes from the presence of the magnetic mass in 
the numerator, while only 14\% decrease comes from the presence of magnetic 
screening in the denominator of the effective crossection. The reason behind 
this is that introduction of magnetic screening in the denominator of 
effective crossection does not regulate the logarithmic divergence, as might 
be expected from Eq.~(\ref{DeltaEDyn_new}). This is because this divergence is 
already naturally regulated in Eq.~(\ref{DeltaEDyn}), where all the relevant 
diagrams are taken into account~\cite{MD_PRC,DH_PRL}. 

\section{Summary}

This paper generalizes dynamical energy loss formalism to non-zero magnetic 
screening. While the introduction of magnetic mass into any perturbative 
calculation is inherently phenomenological, the presented inclusion of the 
effects of modified gluon self energy into our radiative energy loss formalism 
is of general validity as long as a well defined quasiparticle system is 
assumed. Analysis of the finite magnetic mass effects leads to 
a constraint that it is impossible to create a plasma with magnetic mass 
larger than electric. Results presented in this paper allow including 
non-zero magnetic screening into jet suppression calculations, and open a 
possibility for more accurate mapping of QGP properties.

\bigskip
{\it Acknowledgments:}
Valuable discussions with Joseph Kapusta, Antony Rebhan, and Miklos Gyulassy 
are gratefully acknowledged. This work is supported by Marie Curie 
International Reintegration Grant within the $7^{th}$ European Community 
Framework Programme (PIRG08-GA-2010-276913) and by the 
Ministry of Science and Technological Development of the Republic of Serbia,
under projects No. ON171004 and ON173052. Marko Djordjevic is supported in part
by Marie Curie International Reintegration Grant within the $7^{th}$ European 
Community Framework Programme (PIRG08-GA-2010-276996).


\end{document}